\documentclass[letterpaper, 10 pt, conference]{ieeeconf}  
\IEEEoverridecommandlockouts
\usepackage{amsmath, amssymb, amsfonts}
\usepackage{graphicx, textcomp, color, soul}
\usepackage{hyperref, soul, color, comment, makecell}
\usepackage{comment}
\usepackage{tikz}
\usepackage{multirow}

\usepackage{cite}

\def\BibTeX{{\rm B\kern-.05em{\sc i\kern-.025em b}\kern-.08em
    T\kern-.1667em\lower.7ex\hbox{E}\kern-.125emX}}
\begin{document}

\definecolor{color_rev}{rgb}{0, 0, 0}


\title{\LARGE \bf BiND: A Neural Discriminator–Decoder for Accurate Bimanual Trajectory Prediction in Brain–Computer Interfaces }

\author{
Timothée Robert$^{1}$, MohammadAli Shaeri$^{1}$, Mahsa Shoaran$^{1}$%
\thanks{$^{1}$ Institutes of Electrical and Micro Engineering and Neuro-X, EPFL, Lausanne and Geneva, Switzerland.}%
\thanks{This work was supported in part by the Swiss State Secretariat for Education, Research and Innovation (SERI) under the SwissChips initiative, in part by the Swiss National Science Foundation (SNSF) Project Funding 200021\_219752,   in part by the Swiss State Secretariat for Education, Research and Innovation under Contract SCR0548363, and in part by  the WYSS Digital Bridge Project 532932.
}
}

\maketitle

\begin{abstract}
Decoding bimanual hand movements from intracortical recordings remains a critical challenge for brain–computer interfaces (BCIs), due to overlapping neural representations and nonlinear interlimb interactions. We introduce BiND (Bimanual Neural Discriminator–Decoder), a two-stage model that first classifies motion type (unimanual left, unimanual right, or bimanual) and then uses specialized GRU‐based decoders—augmented with a trial-relative time index—to predict continuous 2D hand velocities. We benchmark BiND against six state-of-the‐art models (SVR, XGBoost, FNN, CNN, Transformer, GRU) on a publicly available 13-session intracortical dataset from a tetraplegic patient.
BiND achieves a mean $R^2$ of 0.76 ($\pm$0.01) for unimanual and 0.69 ($\pm$0.03) for bimanual trajectory prediction, surpassing the next-best model (GRU) by +2\% in both tasks. It also demonstrates greater robustness to session variability than all other benchmarked models, with accuracy improvements of up to 4\% compared to  GRU in cross-session analyses.
This highlights the effectiveness of task-aware discrimination and temporal modeling in enhancing bimanual decoding.
\end{abstract}

\section{Introduction}
According to the World Health Organization (WHO), neurological conditions such as stroke and brain injuries affect over one-third of the global population and represent a leading cause of disability \cite{steinmetz2024global, who2025disability}. 
Around 2\% of people worldwide require rehabilitation or assistive technologies~\cite{krahn2011world, WHO2011Disability}, often due to motor impairments from spinal cord injuries, stroke, or related disorders, which can lead to partial or complete paralysis and severely impact quality of life.

To alleviate this burden and restore essential motor functions, particularly hand movements critical for interacting with the environment, intracortical brain-computer interfaces (BCIs) use advanced machine learning algorithms to translate brain-intended actions into movement~\cite{patrick2024state, shaeri2022data}. These systems typically acquire intracortical or ECoG signals from motor-related brain regions, converting intention into control commands for prosthetic devices, robotic limbs, or gait restoration prostheses~\cite{lorach2023walking, ajiboye2017restoration-breif, aflalo2015decoding, flesher2021brain}. The success of BCIs depends on accurately interpreting these neural signals to provide intuitive and seamless motor control for the users.

A wide variety of decoding models have been developed to translate complex neural activity into actionable movement commands.
Classical machine learning approaches, such as linear models and decision trees, have been used for their low computational cost~\cite{YS2024JSSC, Shaeri2024MiBMI, glaser2020machine, an2022power, Shaeri2022Challenges, yao2022fast, shin2022neuraltree, Shoaran2024Intelligent}.
However, given the intricate spatio-temporal dynamics of brain signals, advanced deep learning architectures offer greater representational power. Feedforward neural networks (FNNs) utilize layered nonlinear transformations~\cite{willsey2022real}, while recurrent neural networks (RNNs) are well-suited for handling temporal dependencies~\cite{Willett2021, deo2024brain}. Convolutional neural networks (CNNs) excel at extracting spatial features~\cite{liu2022deep, livezey2021deep}, and hybrid CNN-RNN models further improve performance~\cite{tantawanich2024systematic}. More recently, transformer-based models with attention mechanisms have emerged as powerful tools for efficient sequence modeling and parallel computation, demonstrating potential for neural decoding applications~\cite{vaswani2017attention}.

Bimanual tasks such as eating or dressing are essential for independence but remain poorly supported by current BCIs~\cite{edemekong2019activities, deo2024brain}.
A key challenge is the overlapping neural representations of movements from both hands, often leading to limited decoding accuracy, particularly for the non-dominant hand.
In this paper, we evaluate state-of-the-art neural decoding models for movement trajectory prediction.
To further improve accuracy, we introduce a novel Bimanual Neural Discriminator-Decoder (BiND) model, which incorporates a classification stage to distinguish motion types (unimanual left, unimanual right, or bimanual) before trajectory decoding.
BiND then integrates a GRU-based RNN with a modified recurrent architecture for modeling long-term temporal patterns in addition to short-term dependencies. 
This hybrid model significantly enhances decoding accuracy for both unimanual and bimanual movements, improving the practical viability of BCIs for restoring natural motor function in individuals with paralysis.

\section{Dataset and Proposed Methods}


\begin{figure}[!t]
\centering
{\includegraphics[width=0.49\textwidth]{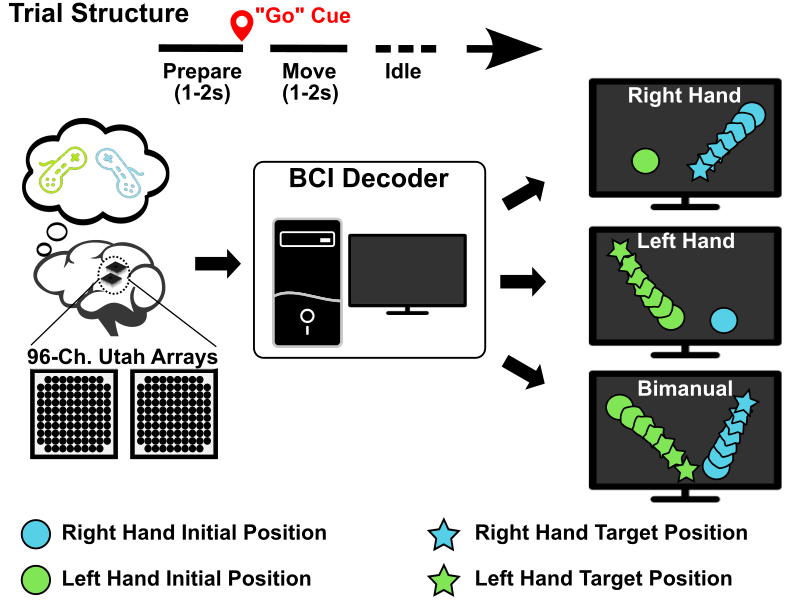}}
\vspace{-.5cm}
\caption{\textbf{Experimental setup}.
The patient faces a screen displaying two cursors (right hand: blue circle; left hand: green circle) and their respective targets (right hand: blue star; left hand: green star). During the 1–2 second ``prepare'' phase, lines connect each cursor to its target, allowing the patient to anticipate the motion. At the ``go'' cue, the cursor(s) automatically move toward their target(s) (one for unimanual, and both for bimanual motion), while the patient imagines controlling them via joystick movement. Each trial ends with a short idle phase before the next cue.
}
\label{experimental_setup}
\vspace{-.30cm}
\end{figure}

\subsection{Task and Data} \label{Datasets}

We used a publicly available dataset introduced in~\cite{deo2024brain}, containing neural recordings from a tetraplegic participant performing a bimanual cursor control task through imagined joysticks movements. During recording sessions, the participant sat facing a screen displaying two pairs of cursor-target markers, one for each hand. Each pair included a movable cursor, representing the participant's intended hand movement, and a stationary target, indicating the aimed position the cursor had to reach. Unimanual trials involved the movement of a single cursor (left or right), while bimanual trials required simultaneous movement of both cursors. The participant was instructed to imagine manipulating joysticks with each hand to guide the cursors toward their respective targets. \textcolor{color_rev}{The dataset was obtained from a pilot clinical trial approved under an Investigational Device Exemption (IDE) by the US Food and Drug Administration, the Stanford University Institutional Review Board, as well as the Mass General Brigham IRB.}

As depicted in Fig.~\ref{experimental_setup}, each trial was structured into three phases: preparation, movement, and idle.
The preparation phase was a brief period during which the participant was cued on the upcoming movement type and direction for motor planning.
In the subsequent movement phase, they performed the imagined movement as the cursor(s) moved toward the target(s). Finally, the idle phase provided a brief rest before the next trial.

The dataset includes recordings from 13 open-loop BCI  sessions, each following the preparation-movement-idle structure. During these sessions, the cursors moved automatically toward their targets while neural activity was recorded alongside the participant’s imagined movement intent.
Neural data were collected from two 96-channel Utah microelectrode arrays implanted in the hand knob area of the left (dominant) precentral gyrus, a motor-related cortical region.
Neural spikes were detected via thresholding~\cite{razmpour2015signal}, i.e., activities exceeding a threshold of 3.5$\times$ the root mean square (RMS) voltage.
These threshold crossing events were then counted in 20-ms intervals. The dataset also includes behavioral measurements, specifically the 2D cursor and target positions at each time step, used to train and evaluate decoding models.

\begin{figure}
\centering
\centering
\includegraphics[width=.75\linewidth]{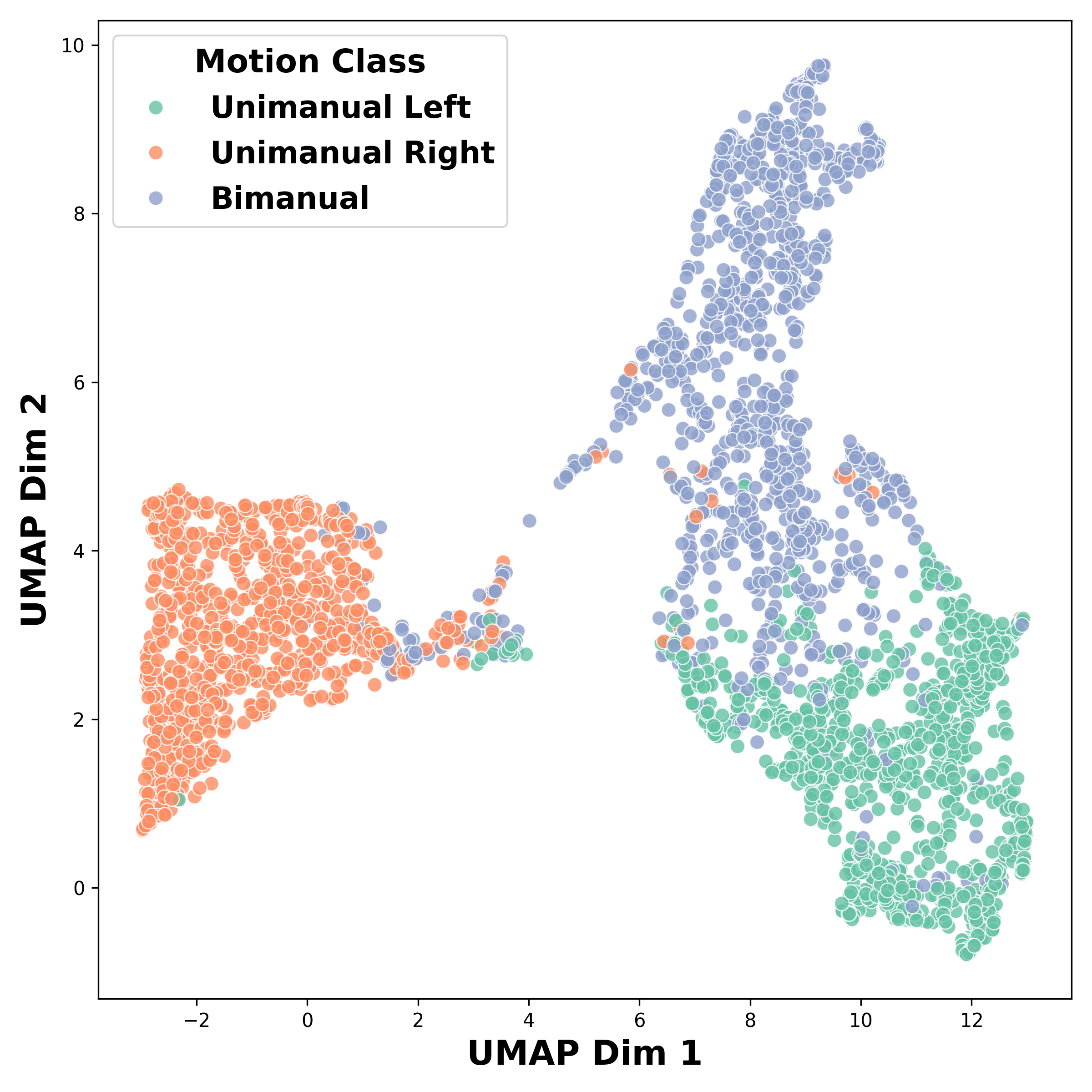}
\vspace{-.25cm}
\caption{\textbf{Latent space visualization}.
The latent space, visualized using Uniform Manifold Approximation and Projection (UMAP), reveals distinct clustering of motion types. Right-hand, left-hand, and bimanual movements are well-separated, with minimal overlap, demonstrating high discriminability between motion classes. Although bimanual trials exhibit partial overlap with unimanual clusters, over 80\% of bimanual data points are correctly classified, indicating strong separability overall.
}
\label{umap}
\vspace{-.30cm}
\end{figure}

\subsection{Data Processing and Training} \label{Data Processing}
Threshold-crossing counts, used as estimates of neural activity, were normalized on a per-session basis and smoothed along the temporal axis using a Gaussian kernel. This reduced temporal noise and variability while preserving rapid signal changes, improving model stability~\cite{sodagar2025real}. 
In our analysis, a kernel width of $\sim$40ms yielded optimal decoding performance across models. 

For each trial, data were extracted from the onset of the ``go'' cue to the end of the ``movement'' phase. 
Following the preparation phase, trial data were segmented into overlapping 600-ms windows (30 time bins) with a 300-ms stride (15 bins), resulting in 50\% temporal overlap between consecutive segments.
Each window served as an independent decoding sample.
The selected window length and stride were empirically chosen to achieve a suitable trade-off between decoding accuracy and temporal resolution.

Among the 13 available sessions, the most recent sessions were designated as the target session, while the preceding sessions were used for training and validation via 10-fold cross-validation. After initial training, 40\% of the target session trials were used to fine-tune the model, allowing it to adapt to inter-session variability. The remaining 60\% were reserved for final evaluation.
By restricting the decoder to past information only (i.e., no access to future time steps), the setup guarantees causal decoding, making the results directly applicable to real-time BCIs.

\begin{figure}
\centering
\includegraphics[width=.99\linewidth]{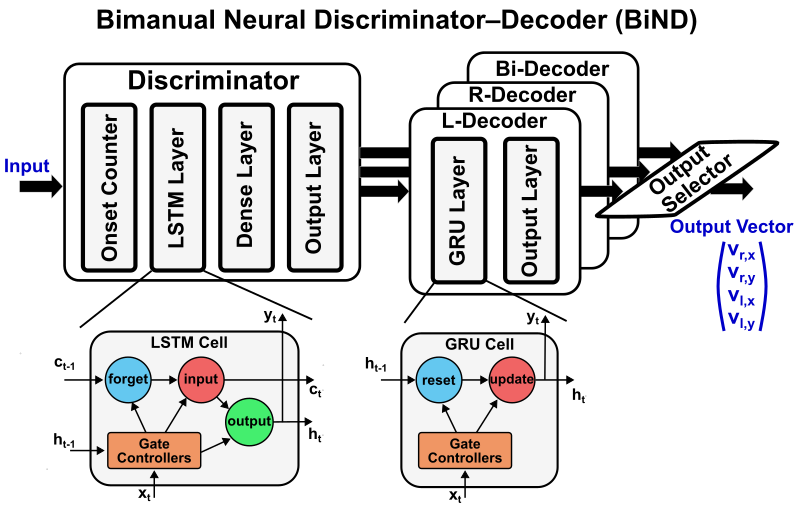}
\vspace{-.60cm}
\caption{
\textbf{BiND model architecture}.
The BiND model processes time-indexed input data \textcolor{color_rev}{with $192\times 30$ dimensions (30 time points, 192 channels)} for enhanced long-term temporal awareness. It consists of a Discriminator that predicts the motion type (right-hand, left-hand, or bimanual) and selects one out of the three decoders tailored to each. 
LSTM and GRU cells are used to capture temporal dependencies, with built-in gating mechanisms to control information flow.
}
\label{fig:unified_pipeline}
\vspace{-.30cm}
\end{figure}

\subsection{The Proposed BiND Model}\label{BiND}
To precisely decode a diverse set of movements from high-dimensional neural signals, we introduce BiND (Bimanual Neural Discriminator-Decoder): a two-stage architecture that explicitly distinguishes between unimanual and bimanual movements and incorporates temporal context via a time-indexing feature (counted from activity onset).
To assess whether these movement types are separable at the neural level, we applied UMAP to the neural data, revealing clear distinctions between left, right, and bimanual movements (Fig.~\ref{umap}).
This structure motivated the BiND architecture, which first classifies the movement type and then routes the signal to specialized decoders, enabling more effective modeling of movement-specific neural dynamics.

An overview of the BiND architecture is shown in Fig.~\ref{fig:unified_pipeline}. The model processes neural input samples to predict 2-dimensional velocities for both hands ($V_{r,x}$, $V_{r,y}$, $V_{l,x}$, $V_{l,y}$). The first stage, the ``Discriminator'', classifies the movement type—unimanual left, unimanual right, or bimanual. It consists of:

\begin{itemize}
\item A 128-unit LSTM layer to capture temporal dependencies in  neural signals,
\item A 64-unit dense layer to refine the representation, and
\item An output layer with a sigmoid activation that generates class probabilities.
\end{itemize}
To mitigate overfitting, a dropout rate of 0.3 was applied to both LSTM and dense layers.

Based on the predicted movement type, each input sample is passed to one of three tailored decoders:
(I) ``L-Decoder'', trained exclusively on unimanual left-hand movements;
(II) ``R-Decoder'', trained exclusively on unimanual right-hand movements; and
(III) ``Bi-Decoder'', trained on all movement types to capture inter-limb coordination and cross-hand interactions.
Interestingly, performance improves when the decoder is trained on all types of data, suggesting that unimanual trajectories help uncover patterns also relevant for bimanual movements. \textcolor{color_rev}{This is consistent with prior findings showing that bimanual motion shares common information with the unimanual context \cite{cross2020maintained}, enabling the decoder to capture these shared components as well as the mixed patterns unique to bimanual motion. Using the discriminator allows the model to learn several simpler mappings rather than a single highly complex one, which could otherwise dilute performance. Furthermore, \cite{deo2024brain} evaluated the tuning properties of different electrodes. In particular, an electrode strongly tuned to one motion type but not another may introduce noise when applied in the latter context.}

All three decoders follow a shared architecture, as follows:
\begin{itemize}
\item A 512-unit GRU layer to reconstruct continuous movement trajectories,
\item A dense output layer that predicts four values: horizontal and vertical cursor velocities for both left and right hands.
\end{itemize}

Moreover, BiND integrates an onset counter that provides a time index, indicating the relative position of segments within the trial.
This auxiliary feature acts as a surrogate for long-range temporal dependencies, enabling the model to recover parts of the lost temporal structure. 
\textcolor{color_rev}{Using GRU layers instead of LSTM  in the decoders showed similar accuracy with better computational efficiency, critical in iBCI applications. The total training time with GRU layers was around 80\% of that with LSTM layers. Nonetheless, LSTMs were retained in the discriminator, as they performed slightly better than GRUs in this role.} 
\textcolor{color_rev}{The discriminator contains roughly 500,000 parameters, whereas a decoder has about 2 million. } 

In summary, BiND enhances neural decoding performance by combining explicit motion-type discrimination, specialized decoding pathways, and modeling of both short- and long-term temporal context.
This offers a principled approach to decoding both unimanual and bimanual motor intentions.

\section{Results}
This section presents the evaluation results for the proposed BiND model, along with comparisons to state-of-the-art neural decoders. We employ the $R^2$ score and Pearson correlation coefficient as the evaluation metrics. The $R^2$ quantifies decoding accuracy by normalizing the reconstruction error with respect to the total variance in the ground truth data~\cite{Kalbasi2024DPARS}. To ensure a fair comparison, all models were trained using the same preprocessing steps and followed identical training pipelines.


\begin{figure}
\centering
\includegraphics[width=\linewidth]{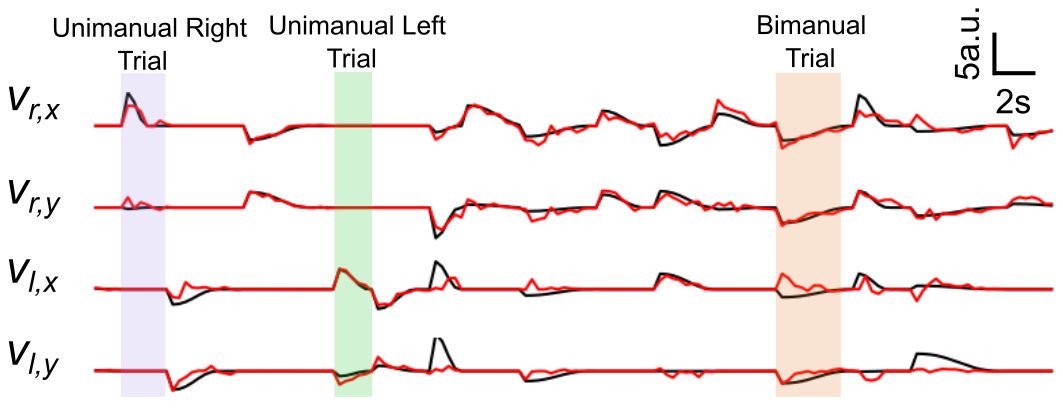}
\vspace{-0.65cm}
\caption{\textbf{Decoded hand velocities using BiND.}  Desired (black) and BiND-predicted (red) velocities along the \(x\)- and \(y\)-axes for right (top) and left (bottom) hands during unimanual and bimanual trials. The model accurately tracks motor intentions, demonstrating generalization to unseen data.}
\label{real_vs_predicted}
\vspace{-0.30cm}
\end{figure}

\subsection{Performance Evaluation} 
The proposed BiND decoder processes neural data in sequential 600-ms windows with 300-ms overlap, spanning from the onset of the ``prepare'' phase to the end of the ``move'' phase.
During inference, it reconstructs continuous hand velocity trajectories for both the left and right hands.
Fig.~\ref{real_vs_predicted} illustrates three typical trials comparing ground-truth (black) and BiND-predicted (red) hand velocity components along the \(x\)- and \(y\)-axes, one for each motion type.
The top rows show right-hand velocities ($V_{r,x}$, $V_{r,y}$), while the bottom rows correspond to left-hand velocities ($V_{l,x}$, $V_{l,y}$).
These plots illustrate BiND's ability to accurately track the temporal dynamics and directional trends of imagined motor outputs on previously unseen data.


\begin{figure}
\centering
\centering
\includegraphics[width=\linewidth]{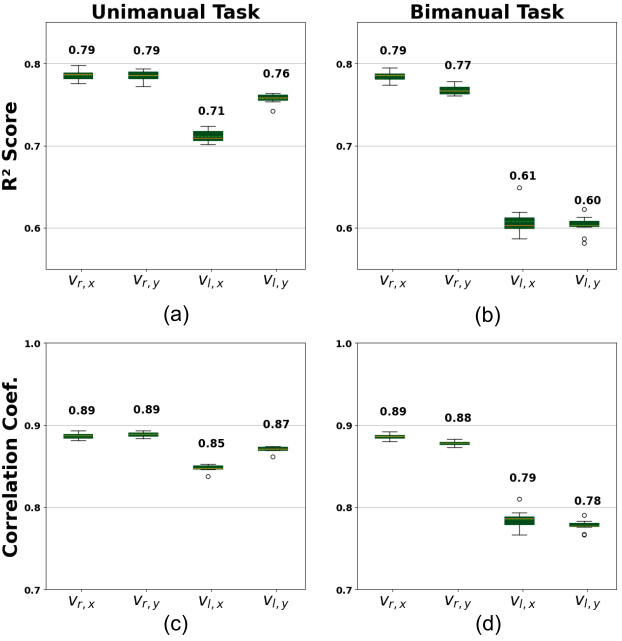}
\vspace{-.65cm}
\caption{\textbf{BiND Decoding Accuracy}.
Box plots show the distribution of $R^2$ scores and correlation coefficients for the four hand velocity components ($V_{r,x}$, $V_{r,y}$, $V_{l,x}$, $V_{l,y}$). 
Panels (a) and (b) display the average $R^2$ scores for unimanual and bimanual cases, respectively, highlighting a disparity in left-hand decoding performance.
The interquartile range (IQR, box) captures the middle 50\% of values (Q1–Q3), with whiskers extending to $\pm 1.5 \times$ IQR. The orange line indicates the mean accuracy, while individual dots represent outliers. 
Correlation analysis in panels (c) and (d) shows reduced variance in bimanual cases, suggesting that $R^2$ is more sensitive to performance variation and better differentiates decoding accuracy across movement types. \textcolor{color_rev}{The strongest outliers observed for the left hand in the bimanual context are likely attributable to inter-session neural variability combined with the added complexity of decoding the non-dominant hand. } 
}
\label{baseline_comparison}
\vspace{-.30cm}
\end{figure}

Fig.~\ref{baseline_comparison}(a) shows the decoding accuracy of 2-dimensional hand velocity components
during unimanual movements. 
BiND achieves an \(R^2\) score of 0.79 for the right hand (95\% CI: $[0.781, 0.790]$ for both \(V_{r,x}\) and \(V_{r,y}\)) and 0.71–0.76 for the left hand (95\% CI: $[0.706, 0.717]$ for \(V_{l,x}\), $[0.753, 0.762]$ for \(V_{l,y}\)), indicating strong reconstruction of single-hand trajectories.
A similar trend appears in the correlation coefficient (Fig.~\ref{baseline_comparison}(c)), which also reflects high agreement between predicted and ground-truth signals.

Fig.~\ref{baseline_comparison}(b)\&(d) reveals a notable disparity between left and right hand movements during bimanual tasks.
While right-hand decoding remains high, left-hand accuracy drops by 10–16\% in \(R^2\) and 6–9\% in correlation.
This degradation suggests that decoding the non‐dominant hand becomes more challenging during coordinated bilateral tasks.
These findings align with prior works~\cite{aramaki2006suppression, tzourio2015between, deo2024brain}, 
which reported suppressed neural tuning for the ipsilateral (left) hand and relatively stable tuning for the contralateral (right) hand in similar contexts.
Together, these results underscore the value of BiND architecture that explicitly separates decoding pathways and accounts for asymmetric neural encoding of bilateral motor control.

It is important to note that the correlation coefficient tends to emphasize high-variance segments (typically high-velocity movements).
In contrast, the \(R^2\) score measures the proportion of total variance explained by the model and penalizes errors uniformly across all velocity magnitudes.
Therefore, \(R^2\) offers a more balanced assessment of performance across both fast and slow movements, whereas the correlation coefficient can overestimate performance if fine or low-amplitude movements are underrepresented.

\begin{figure}
\centering
\includegraphics[width=\linewidth]{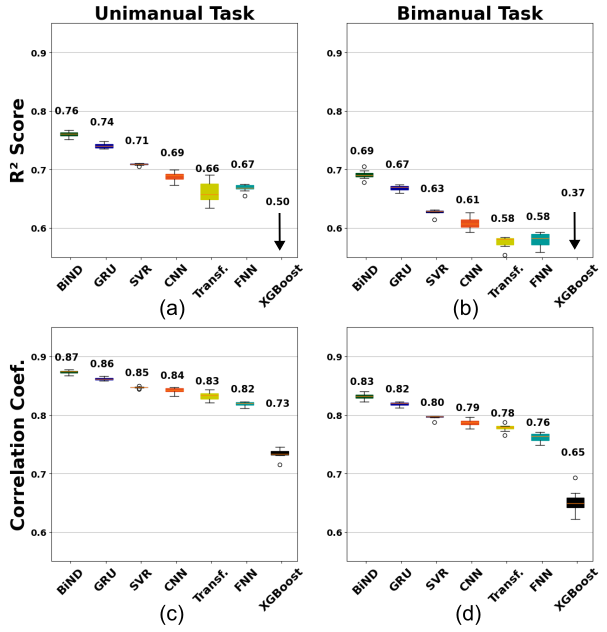}
\vspace{-.65cm}
\caption{
\textbf{Decoding performance comparison across models.} 
Unimanual and bimanual decoding results are reported as $R^2$ scores (a--b) and Pearson correlation coefficients (c–d), averaged across cross-validation folds. BiND significantly outperforms all baseline models, exceeding 0.76 $R^2$ in the unimanual task (a) and 0.69 in the bimanual task (b). It also surpasses 0.87 correlation in the unimanual case (c) and 0.83 in the bimanual case (d). Compared to the next-best model (GRU), BiND improves $R^2$ by approximately 2\% and correlation by 1\%.
}
\label{all_models}
\vspace{-.30cm}
\end{figure}

\subsection{Comparison with State-of-the-Art Models}
We implemented and benchmarked seven state-of-the-art and conventional architectures for movement decoding: Support Vector Regression (SVR), XGBoost, Feedforward Neural Network (FNN), Gated Recurrent Unit (GRU), Convolutional Neural Network (CNN), Transformer (as employed in prior studies~\cite{glaser2020machine, Shaeri2022Challenges, deo2024brain, Kalbasi2024DPARS, shaeri2025integrating}), and the proposed BiND model. Each model was trained to predict trial-averaged hand velocities in the $x$- and $y$-directions for both the right and left hands.


Figure~\ref{all_models} presents the decoding accuracy of all models for both unimanual and bimanual tasks.
BiND consistently outperforms the alternatives in both scenarios. It is the only model to achieve an \( R^2 \) score exceeding 0.76 in the unimanual task (Fig.~\ref{all_models}(a)) and reaches 0.69 in the more challenging bimanual setting (Fig.~\ref{all_models}(b)).
The performance drop of $\sim$7\% when transitioning from unimanual to bimanual decoding reflects the increased complexity of decoding two-hand movements, especially for the non-dominant hand.
A similar trend is observed in the correlation coefficients (Fig.~\ref{all_models}(c)--(d)), where BiND remains the top performer, reaching 0.87 and 0.83 in the unimanual and bimanual tasks, respectively. \textcolor{color_rev}{The results highlight the asymmetric nature of neural representations for bimanual motion, consistent with benchmarked models and prior literature. Nevertheless, neural plasticity has been shown to alter these representations, promoting greater symmetry through bimanual training, as observed in musicians\cite{takeuchi2012motor}.}

BiND surpasses GRU, our second-best model, by 2\% in terms of \( R^2 \) score for both unimanual and bimanual decoding tasks.
This result underscores the strength of RNN-based architectures (BiND and GRU), which consistently yield superior decoding performance.
In general, RNN-based models outperform non-recurrent alternatives by 3--7\% in \( R^2 \) (Figs.~\ref{all_models}(a) and~\ref{all_models}(b)) and by 1--5\% in correlation coefficient (Figs.~\ref{all_models}(c) and~\ref{all_models}(d)).
These findings validate the choice of an RNN backbone in BiND, particularly for capturing the short- to mid-range temporal dependencies~\cite{johnston2025revisiting} inherent in movement trajectory decoding.

CNN, which primarily models localized temporal dynamics, achieves moderate performance ($R^2$ of 0.71 in unimanual and 0.61 in bimanual decoding), but falls short of the RNN-based models. CNNs rely on local connectivity and receptive fields~\cite{wang2018non, shaeri2025integrating}, which may limit their ability to capture the long-range inter-regional dependencies characteristic of brain activity. This suggests that temporal mechanisms are more critical than local spatial patterns for decoding neural signals in this task.
The Transformer, despite its strength in capturing long-term temporal dependencies, performs relatively poorly on our dataset ($R^2$ of 0.66 in unimanual, 0.58 in bimanual decoding).
While surprising at first glance, given the model’s success in modeling patterns in sequential data~\cite{wen2022transformers}, this result suggests that short- and mid-term temporal dynamics are more critical for decoding hand trajectories from neural signals than long-range dependencies. \textcolor{color_rev}{Indeed, Transformers are known to better handle long-range dependencies, and the 30 bins time windows used in this project may be suboptimal for such models. Furthermore, the use of non-overlapping windows can hinder  Transformer performance, as each window is processed independently, whereas RNNs can benefit from this setup due to their inherent memory mechanisms.}

Similarly, FNN yields relatively low performance ($R^2$ of 0.67 in unimanual, 0.58 in bimanual decoding), likely due to its inability to model temporal structures~\cite{glaser2020machine, zhang2015feedforward}, which are essential for capturing the dynamic nature of neural signals underlying continuous movement trajectories.
SVR with an RBF kernel performs moderately well ($R^2$ of 0.71 in unimanual and 0.63 in bimanual decoding), likely due to its ability to independently model each output using a standard kernel shape~\cite{tran2024critical}, allowing it to effectively capture simpler, low-dimensional mappings between neural features and kinematics.
Finally, XGBoost is by far the worst-performing model. It achieves an $R^2$ score of 0.50 in the unimanual case—lower than the FNN's performance in the bimanual setting—and drops to just 0.37 in the bimanual task, a 13\% decline. This is expected, as XGBoost cannot model temporal dependencies, and its ability to capture complex nonlinearities is more limited than the deep learning models evaluated in this work. Furthermore, XGBoost requires the neural data to be flattened, leading to additional information loss.

In the above comparisons, we evaluated all models on the final session, using all preceding sessions for training to ensure effective utilization of the available data. To further assess cross-session generalizability, we tested the models on the last three sessions, which included both unimanual and bimanual movements. The results demonstrated that model performance improved with an increasing number of training sessions. When evaluated across all sessions, BiND achieved accuracies of 0.72 for unimanual and 0.6 for bimanual decoding, while GRU reached 0.7 and 0.56, respectively. In all cases, BiND consistently outperformed GRU, showing a 2–4\% improvement in accuracy and indicating greater adaptability and robustness to session variability. \textcolor{color_rev}{Ablation studies revealed that both components introduced in this work contributed comparably.} Notably, non-recurrent models exhibited a marked decline in performance, underscoring the critical role of temporal modeling in achieving reliable neural decoding across sessions.


Overall, BiND’s architecture comprises separate decoding pathways for left, right, and bimanual trajectories, enabled by the neural discriminator.
It also leverages a hybrid 
recurrent structure to effectively capture both short-term and long-term temporal dependencies, aided by time indexing relative to movement onset. \textcolor{color_rev}{The model exhibits an inference time of 100-150ms on GPU, largely due to its recurrent nature. However, optimization techniques can substantially reduce both inference time and model size while maintaining similar accuracy.}

\section{Conclusion} \label{Conclusion}

In this work, we presented BiND, a novel bimanual decoding algorithm that integrates a motion‐type discriminator with specialized GRU‐based decoders and a time‐index feature. 
BiND consistently outperformed six state‐of‐the‐art baselines—including SVR, XGBoost, CNN, FNN, Transformer, and standard RNNs. 
These results demonstrate that task‐aware specialization (discriminating unimanual vs. bimanual movements) and explicit temporal modeling (short‐ and long‐range dependencies via GRUs and time indexing) are critical for high‐fidelity two‐hand trajectory prediction. Importantly, our causal decoding pipeline—validated under inter‐session fine‐tuning—ensures robustness to session variability and is directly applicable to real‐time BCI implementations.  \textcolor{color_rev}{Future work will extend BiND to adaptive online settings across multiple participants, explore lightweight implementations for embedded BCI hardware, and investigate integration with sensory feedback channels. BiND will also be assessed on other types of tasks, such as reaching and grasping with robotic prosthetic hands, especially in the challenging context of bimanual coordination. Adapted to real-life applications, BiND has the potential to establish a new benchmark for impaired patients, providing simpler and more intuitive hand control.}
By advancing accurate bimanual decoding, BiND brings BCIs one step closer to restoring coordinated, naturalistic hand function for paralyzed patients.

\end{document}